\def\Version{{5
    }}




\message{<< Assuming 8.5" x 11" paper >>}    

\magnification=\magstep1	          
\raggedbottom

\parskip=9pt

%

\def\singlespace{\baselineskip=12pt}      
\def\sesquispace{\baselineskip=16pt}      







\font\openface=msbm10 at10pt
 %

\def\Integers      {{\hbox{\openface Z}}}

\def\Reals         {{\hbox{\openface R}}}

 %
 %
 %



\font\german=eufm10 at 10pt

\def\Buchstabe#1{{\hbox{\german #1}}}










%

%
%



\def\implies{\Rightarrow}

%



\def\sqr#1#2{\vcenter{
  \hrule height.#2pt 
  \hbox{\vrule width.#2pt height#1pt 
        \kern#1pt 
        \vrule width.#2pt}
  \hrule height.#2pt}}




\def\lto{\mathop
        {\hbox{${\lower3.8pt\hbox{$<$}}\atop{\raise0.2pt\hbox{$\sim$}}$}}}
\def\gto{\mathop
        {\hbox{${\lower3.8pt\hbox{$>$}}\atop{\raise0.2pt\hbox{$\sim$}}$}}}
%
%
%



\def\part{\subseteq}		


\def\braces#1{ \{ #1 \} }



\def\to{\mathop\rightarrow}	


\def\less{\backslash}		


\def\interior #1 {  \buildrel\circ\over  #1}     




\def\basisvector#1#2#3{
 \lower6pt\hbox{
  ${\buildrel{\displaystyle #1}\over{\scriptscriptstyle(#2)}}$}^#3}



\def\hat{\widehat}		



\fontdimen16\textfont2=2.5pt
\fontdimen17\textfont2=2.5pt
\fontdimen14\textfont2=4.5pt
\fontdimen13\textfont2=4.5pt




%



%
%
%
%
%
%
%
%
%
%
%
%
%
%
%

%
 \let\miguu=\footnote
 \def\footnote#1#2{{$\,$\parindent=9pt\baselineskip=13pt%
 \miguu{#1}{#2\vskip -7truept}}}
%
%

\def\linebreak{\hfil\break}
\def\lbr{\linebreak}
\def\pagebreak{\vfil\break}

\def\noparagraph{\vskip -\parskip \smallskip \noindent}

\def\BulletItem #1 {\item{$\bullet$}{#1 }}
\def\bulletitem #1 {\BulletItem{#1}}

\def\THEOREM{\noindent {\csmc Theorem \ }}

\def\PrintVersionNumber{
 \vskip -1 true in \medskip 
 \rightline{version \Version} 
 \vskip 0.3 true in \bigskip \bigskip}

\def\author#1 {\medskip\centerline{\it #1}\bigskip}

\def\address#1{\centerline{\it #1}\smallskip}

\def\furtheraddress#1{\centerline{\it and}\smallskip\centerline{\it #1}\smallskip}

\def\email#1{\smallskip\centerline{\it address for email: #1}} 

\def\AbstractBegins
{
 \singlespace                                        
 \bigskip\leftskip=1.5truecm\rightskip=1.5truecm     
 \centerline{\bf Abstract}
 \smallskip
 \noindent	
 } 
\def\AbstractEnds
{
 \bigskip\leftskip=0truecm\rightskip=0truecm       
 }

\def\section #1 {\bigskip\noindent{\headingfont #1 }\par\nobreak\noindent}

\def\subsection #1 {\medskip\noindent{\subheadfont #1 }\par\nobreak\noindent}
 %

\def\ReferencesBegin
{
 \singlespace					   
 \vskip 0.5truein
 \centerline           {\bf References}
 \par\nobreak
 \medskip
 \noindent
 \parindent=2pt
 \parskip=6pt			
 }
 %

\def\reference{\hangindent=1pc\hangafter=1} 

\def\ref{\reference}

\def\sepref{\parskip=4pt \par \hangindent=1pc\hangafter=0}
 %

\def\journaldata#1#2#3#4{{\it #1\/}\phantom{--}{\bf #2$\,$:} $\!$#3 (#4)}
 %

\def\eprint#1{{\tt #1}}

\def\arxiv#1{\hbox{\tt http://arXiv.org/abs/#1}}
 %

\def\webtilde{\lower2pt\hbox{${\widetilde{\phantom{m}}}$}}

\def\webhome{{\tt http://www.perimeterinstitute.ca/personal/rsorkin/}}
 %

 %



\def\hpf#1{\webhome{\tt{some.papers/}}}
 %

\def\hpfll#1{\webhome{\tt{lisp.library/}}}
 %




\font\titlefont=cmb10 scaled\magstep2 

\font\headingfont=cmb10 at 12pt
%

\font\subheadfont=cmssi10 scaled\magstep1 
%


\font\csmc=cmcsc10  





\def\EA{\Buchstabe{A}}
\def\Z2{\Integers_2}
\def\negate{\neg}		

\def\Ahat{\hat{\Buchstabe{A}}}

\input epsf
\epsfverbosetrue

%
%
%
%
%
 \def\FigureNumberCaption#1#2#3{\vbox{
 \centerline{\vbox{\epsfbox{#1}}}
 \leftskip=1.5truecm\rightskip=1.5truecm     
 \singlespace
 \noindent{{\it Figure #2}. #3}
 \vskip .25in\leftskip=0truecm\rightskip=0truecm}
 \sesquispace}



\phantom{}


\PrintVersionNumber



\sesquispace
\centerline{{\titlefont Logic is to the quantum as geometry is to gravity}\footnote{$^{^{\displaystyle\star}}$}%
{To appear in G.F.R. Ellis, J. Murugan and A. Weltman (eds),
 {\it Foundations of Space and Time} (Cambridge University Press)
}}

\bigskip


\singlespace			        

\author{Rafael D. Sorkin}
\address
 {Perimeter Institute, 31 Caroline Street North, Waterloo ON, N2L 2Y5 Canada}
\furtheraddress
 {Department of Physics, Syracuse University, Syracuse, NY 13244-1130, U.S.A.}
\email{sorkin@physics.syr.edu}

\AbstractBegins                              
I will propose that the reality to which the quantum formalism
implicitly refers is a kind of generalized history, the word history
having here the same meaning as in the phrase sum-over-histories.  This
proposal confers a certain independence on the concept of event, and it
modifies the rules of inference concerning events in order to resolve a
contradiction between the idea of reality as a single history and the
principle that events of zero measure cannot happen (the Kochen-Specker
paradox being a classic expression of this contradiction).  The
so-called measurement problem is then solved if macroscopic events
satisfy classical rules of inference, and this can in principle be
decided by a calculation.  The resulting conception of reality involves
neither multiple worlds nor external observers.  It is therefore
suitable for quantum gravity in general and causal sets in particular.
\AbstractEnds


\sesquispace

\section{1.~Quantum gravity and quantal reality}
Why, in our attempts to unify our theories of gravity and the quantum,
has progress been so slow?  One reason, no doubt is that it's simply a
very hard problem.  Another is that we lack clear guidance from
experiments or astronomical observations.  But I believe that a third
thing holding us back is that we haven't learned how to think clearly
about the quantum world in itself, without reference to  ``observers" and
other external agents.

Because of this we don't really know how to think about the Planckian
regime where quantum gravity is expected to be most relevant.  We don't
know how to think about the vacuum on small scales, or about the inside
of a black hole, or about the early universe.  Nor do we have a way to
pose questions about relativistic causality in such situations.  This is
particularly troubling for the causal set program [1], 
within which a
condition of ``Bell causality" has been defined in the classical case,
and has led there to a natural family of dynamical laws (those of the
CSG or ``classical sequential growth'' models) [2].  
If we possessed an analogous concept of ``quantal Bell
causality", we could set about deriving a dynamics of quantal sequential
growth.  But without an observer-free notion of reality, how does one
give meaning to superluminal causation or its absence in a causal set?

It's not that individual physicists have no notion of what the quantal world is like, of
course.  We all employ intuitive pictures in our work, and for example,
I imagine that very few people think of a rotons in a superfluid in terms
of selfadjoint operators.  But what we lack is a coherent descriptive
framework.  We lack, in other words, an answer to the question, What is
a quantal reality?

My main purposes in this talk are first
to propose an answer to this question (or really a family possible answers),
and second 
to explain how, on the basis of this answer, 
the so called measurement problem can be posed and plausibly solved.
My proposal belongs to the histories-based way of thinking about
dynamics, which in a quantal context corresponds to path-integral
formulations.  More specifically it rests on three or four basic ideas,
that of {\it event}, that of {\it preclusion}, and that of {\it
anhomomorphic inference} concerning {\it coevents}, whose meaning I will
try to clarify in what follows.

\section{2.~Histories and events (the kinematic input)}            
In classical physics, it was easy to say what a possible reality was,
although the form of the answer was not static, but changed as our
knowledge of nature grew. 
Electromagnetic theory, for example,
conceived reality as a background Minkowski spacetime inhabited by a
Faraday field $F_{ab}$ together with a collection of particle
worldlines, each with a given charge and mass, while
reality for General Relativity was a 4-geometry together
with possible matter fields, thus a diffeomorphism-equivalence class of
Lorentzian metrics and other fields. 
Of course, we were far from knowing all the details of the actual
reality, but we could say exactly what in principle it would have taken
to describe reality fully if we did have the details.  Thus we could
survey all the kinematically possible realities, and then go on to state
the dynamical laws (equations of motion or field equations) that further
circumscribed these possibilities.

Another example comes from Brownian motion, which in important ways
stands closer to quantum mechanics than deterministic classical theories
do.  Here, if we imagine that nothing exists but one Brownian particle,
then
a possible reality is just a single worldline (continuous but not
differentiable), and the dynamical law is a set of transition
probabilities, or more correctly a probability-measure on the space of
all worldlines.

Such a possible reality --- a spacetime, a field, a worldline, etc. ---
is what is meant by the word ``history'' in the title of this section;
and according to the view I am proposing, such histories furnish the raw
material from which reality is constructed.\footnote{$^\star$}
{The word history thus denotes the same thing it does when people call
 the path integral a ``sum-over-histories".  To avoid confusion with
 other uses of the word, one might say {\it proto-history} or
 perhaps ``kinematical'' or ``bare'' history.  Given this distinction,
 one might then refer to quantal reality as a ``quantal history''.}
However, unlike in classical physics, we will not (in general) identify quantal
reality with a single history, instead we will have certain sorts of
``logical combinations'' of histories which will be described by 
{\it coevents}.  In the simplest case a coevent will
correspond merely to a set of histories.  Although, without further
preparation, I cannot yet make precise what quantal reality will be, let
me stress at the outset what it will {\it not} be, namely a
wave-function or state-vector.  Nor will the Schr{\"o}dinger equation
enter the story at all.  Such objects can have a technical role to play,
but at no stage will they enter the basic interpretive framework.

As already indicated the concepts of event and coevent will be
fundamental to this framework.  In order to define them, we need first
to introduce the {\it history space} $\Omega$ whose elements are the
individual histories.  An {\it event} is then a subset of $\Omega$.
When $\Omega$ contains an infinite number of histories, not every subset
will be an event because some sort of ``measurability'' condition will
be required, but for present purposes I will always assume that
$|\Omega|<\infty$.  In that case, one can equate the word ``event'' to
the phrase ``subset of $\Omega$''.

Notice that this definition of the word ``event'' parallels its use in
everyday speech, where a sentence like ``It rained all day yesterday''
denotes in effect a large number of more detailed specifications of the
weather, all lumped together under the heading ``rain''.
(This usage of ``event'' also follows the customary terminology in 
probability theory, where however $\Omega$ is often
called the ``sample space'' rather than the history space.)  
On the other hand, one should not confuse event in this sense with the
word ``event'' used to denote a point of spacetime.  The latter may also
be regarded as a type of event, perhaps, but that would only be a very
special case of what I mean by event herein, and it would be relevant
only in connection with quantum gravity per se.

Let us write $\EA$ for the space of all events.  Structurally, $\EA$ is
a Boolean algebra, meaning that union, intersection, complementation and
symmetric-difference are defined for it.  In logical terms these
correspond respectively to the connectives {\it or}, {\it and}, {\it
not}, and {\it xor}.

Finally, we need to define {\it coevent}.  We have defined the dual object, an
event $E\in\EA$, as a subset of $\Omega$, but we can also think of 
an event as
a {\it question}, in our previous example the question 
``Did it rain all day yesterday''?   
A coevent can then be thought of
as something that answers all conceivable questions.  
More formally, a
{\it coevent} will be a map $\phi:\EA\to\Z2$, where $\Z2$ is the
two-element set $\braces{0,1}$,    the
intended meaning being that $\phi(E)=1$ if and only if the event $E$
actually happens.  Thus, $1$ represents the answer ``yes'' (or ``true'')
and $0$ the answer ``no'' (or ``false'').  I will take the point of view
that such a coevent is a full description of reality, quantal or
classical.  After all, what more could one ask for in the way of description than an
answer to every question that one might ask about the world?  
For now we place no conditions on
$\phi$ other than that it takes events to members of $\Z2$.
(Notice that a coevent is in some sense a ``higher order object''.  If
 one thought of events as ``predicates'' then a coevent would be a
 ``predicate of predicates''.\footnote{$^\dagger$}
{A function from a Boolean algebra to $\Z2$ is sometimes called a
 ``truth valuation'', but that terminology would be misleading here.  It
 would suggest too strongly that the event-algebra belongs to some 
 {\it a priori} formal language or logical scheme, with reality being
 merely a ``model'' of that logic.  
 But in the context of physics, 
 it would seem that the events come first and the descriptive language second.
 Moreover, one of the main points of this paper philosophically is (as
 its title indicates) that the rules of logical inference are part of
 physics and can never be written down fully without some knowledge of
 the dynamics.})

\section{3.~Preclusion and the quantal measure (the dynamical input)}
I've said that anhomomorphic logic grows out of the path integral, but
in order to understand what this means, you must think of the path
integral as something more than just a propagator from one wave function
$\psi$ to another.  What, in fact, does the path integral really
compute, if we try to understand it on its own terms?  

Let us for a (very brief) moment adopt an ``operational'' point of view
which only cares about the probabilities of instrumental ``pointer
events''.  Such events can be idealized as ``position measurements''
(the positions of the pointers), and it has been known for a long time
that the joint probabilities for successive position measurements can be
computed directly from the path integral without reference to any wave
function, except possibly as a shortcut to specifying initial conditions.
The probabilities in question here are those furnished by the standard
evolve$-$collapse$-$evolve algorithm to be found in any textbook of quantum
mechanics.  When one abstracts from the mathematical machinery used to
compute them, what remains is a probability measure
$\mu$ on a space of ``instrument events'', and it is this measure,
rather than any wave function, that has direct operational meaning!  How
one may compute $\mu$ directly from the path integral is described in
more detail in reference [3], but for us here the important
points are three.  First that $\mu$ refers not to ``measurements'' per
se, but merely to certain macroscopic happenings, and second that it is
natural, when $\mu$ is expressed as a path-integral, to regard
these macroscopic happenings as being {\it events} in precisely the
sense defined above.  (The histories in this case would specify the
trajectories of the molecules comprising the ``pointer'', and an event
would be, as always, a set of histories.)  
The crucial observation then is that the path integral computation of
$\mu$ makes sense for {\it any} set of histories --- any event --- and
therefore need not be tied to some undefined notion of ``measurement''.

This event-function $\mu:\EA\to\Reals^{+}$ I will call the {\it quantal
measure}, and I will take the point of view that it is the answer to my
question above about what the path integral really computes when we try
to understand it on its own terms.  From this histories vantage point, the
textbook rules for computing probabilities are not fundamental
principles, but rather rules of thumb whose practical success for
certain macroscopic events needs to be explained on the basis of a
deeper understanding of what the quantal measure is telling us about
microscopic reality.  Or to put the point another way,
quantum mechanics formulated via the path integral presents itself to us
as a generalized probability theory with $\mu$ appearing as a
generalized probability measure
[4] [5][6].
Our first task then is to interpret
this generalized measure.  
(Much more could be said about the formal properties of $\mu$ and their
relation to the hierarchy of [7] and to the {\it decoherence
functional} that was first defined within the ``decoherent histories''
interpretation of quantum mechanics [8].  I will, however,
limit myself to these two references and to a reference to an experiment
currently testing the ``3-slit sum rule'' that expresses in great
generality the quadratic nature of the Born rule [9].)
 
One's first thought might be to interpret $\mu(E)$ as an ordinary
probability attaching to the event $E$, but this idea founders at once
on the failure of $\mu$ to be additive on disjoint events.  Since such
non-additivity expresses the physical phenomenon of {\it interference}
that lies at the heart of quantum mechanics, the impasse seems to me to
be definitive:  some other concept than probability in the sense of
relative frequency seems to be called for.  Moreover, $\mu$ fails to
be bounded above by 1, whence some events $E$ would have to be ``more
than certain'', were we to take $\mu(E)$ as a probability in the
ordinary sense.  There is however, one special case in which
normalization and additivity become irrelevant, namely for events $E$
such that $\mu(E)=0$.  In such a case, one could conclude on almost any
interpretation that the event $E$ should never happen.  
(Classically, $\mu(E)$ can never vanish exactly except in trivial cases,
 but quantally it can, thanks precisely to interference!)
Such an event (of $\mu$-measure 0),
I will denote as {\it precluded}, and I propose to interpret $\mu$ in
terms of the following {\it preclusion postulate}:  If $\mu(E)=0$ then
$E$ does not happen.

With respect to a given coevent $\phi$, the ``not happening'' of $E$ is
expressed, as we have seen, by the equation $\phi(E)=0$, and the 
preclusion postulate becomes thereby a rule limiting the coevents
that are dynamically possible:
$$
            \mu(E) =0 \quad \implies \quad \phi(E)=0
$$
A coevent that fulfills this condition, I will call {\it preclusive}.

By isolating in this manner what one might call the purely logical
implications of the generalized measure $\mu$ , one may hope to bring out those
aspects which are peculiarly quantal, as opposed to aspects pertaining to
probability more generally.
Of course it will be necessary at some stage to recover not only the
``logical'' but also the properly probabilistic predictions one obtains from
the standard quantum apparatus.  Whether or not the preclusion
rule above will suffice for this is not entirely clear, but if it does, one will
have cleared up some of the confusion surrounding even the classical
probability concept.  If on the other hand, one needed something more
than the strict preclusion rule,
one could simply extend it 
to embrace
the case of
 ``approximate preclusion'',
where $\mu(E)$ is not exactly zero but still small enough to be treated
as if it vanished.  In this way, the difficulties of classical
probability would not have grown any better, but (hopefully) they would not
have grown any worse either [10].
By basing the probability concept on approximate preclusion, one would
in effect be adopting the interpretation sometimes known as
Cournot's principle, according to which the assertion that an event of
sufficiently small measure will not happen exhausts the scientific
meaning of probability.  (See [11] for a concise
statement of this idea.)\footnote{$^\flat$}
{Predictions about frequencies follow when one construes multiple
 repetitions of some experiment as a single, combined experiment
 grouping all the repetitions together into a single sample space.  The
 event that the overall frequencies come out wrong will then possess a
 tiny measure.}
Cournot's principle is not free of problems, of course,
but neither is any other account of probability, as far as I know.  In any case, it seems
prudent to leave probability aside at first, and concentrate on
the purely logical questions raised by the preclusion principle.  
Considering that
the latter
seem to require a radical revision of some basic logical presuppositions,
questions of probability might appear in a very different light,
once a more adequate picture of quantal reality is in place.

To summarize the burden of this section then,
the idea is that the whole dynamical content of the quantal formalism
reduces to the preclusion rule stated above (possibly supplemented by
its generalization to the case of approximate preclusion).

\section{4.~The three-slit paradox and its cognates}
Viewed through the lens of the path-integral, quantum theory appears as
a generalized theory of stochastic processes characterized by the
quantal measure $\mu$, and this makes feasible a ``histories based'' way
of thinking about the dynamics that seems more suited to the needs
of quantum gravity than alternative accounts inspired by either the
$S$-matrix or the Schr{\"o}dinger equation.
For such an approach to
succeed, however, one needs to free the path integral from its {\it
conceptual} dependence on objects like the wave function.  That is, one
needs a {\it free-standing} histories-based formulation of
quantum theory.  A priori, such a formulation need not base itself on
the path integral, but as things stand, no other alternative has so far
offered itself.  In practice then we can (for now at least) vindicate
the histories-based viewpoint (also called the ``spacetime'' viewpoint)
only by clarifying the physical meaning of the quantal measure $\mu$.

I have proposed above that the dynamical implications of $\mu$ are
mediated by what it tells us about the precluded events, the sets of
histories of zero measure in $\Omega$.  Perhaps there is more to it than
this, but even if preclusion is not the full story, it is hard to see
how --- without entirely abandoning the attempt to interpret $\mu$ as
some sort of generalized probability measure --- one could avoid the
implication that events of measure zero do not happen.  If this is
correct, then acceptance of the preclusion principle is a minimal
requirement for re-conceiving quantum mechanics along lines suggested by
the path-integral formalism.

The problem then is that, thanks to interference, there are far too many
sets of measure zero, so many in fact that events which are in reality
able to occur seem to be ruled out as a logical consequence of the
preclusion of other events that overlap them.  (Remember that event =
subset of $\Omega$.)  Here I'm referring to the numerous ``logical
paradoxes'' of quantum theory, including the Kochen-Specker
paradox,\footnote{$^\star$}
{This comes in two versions, the original version referring to a single
 spin-$1$ system, and the version of Allen Stairs [12] referring to an
 entangled pair of such systems.  The latter seems to have been the
 first example of an obstruction to locally causal theories based purely
 on logic (as opposed to probability-based obstructions like the Bell
 inequalities).} 
the GHZ paradox, the Hardy paradox, and the ``three-slit paradox'' that
I'll focus on in a moment.  Each of these can be realized in terms of
sets of particle trajectories together with appropriate combinations of
slits or Stern-Gerlach-like devices 
(as in [13] or [14] for example), 
so that the relevant quantal
measure can be discerned.
What then makes all
these paradoxes paradoxical is that all or part of the history space
$\Omega$ is covered by precluded events.  In the Kochen-Specker setup,
these overlapping preclusions cover the whole of $\Omega$, implying, according
to our customary way of reasoning, that nothing at all can happen
(cf. [15]).
The other examples are similar, but not quite as dramatic.

The contradictions in question
can be illustrated with a diffraction experiment involving not two, but
three ``slits''.
Consider then, an idealized arrangement as shown, with source $S$,
apertures $a$, $b$, $c$, and a designated location $d$ to which the
particle in question might or might not travel.  (The letter $d$ is
meant to suggest ``detector'', but modeling one explicitly would
complicate our setup unnecessarily, without changing anything essential,
as long as we can assume that the detector would function properly.)

\epsfxsize=3.6in
\bigskip
\FigureNumberCaption
{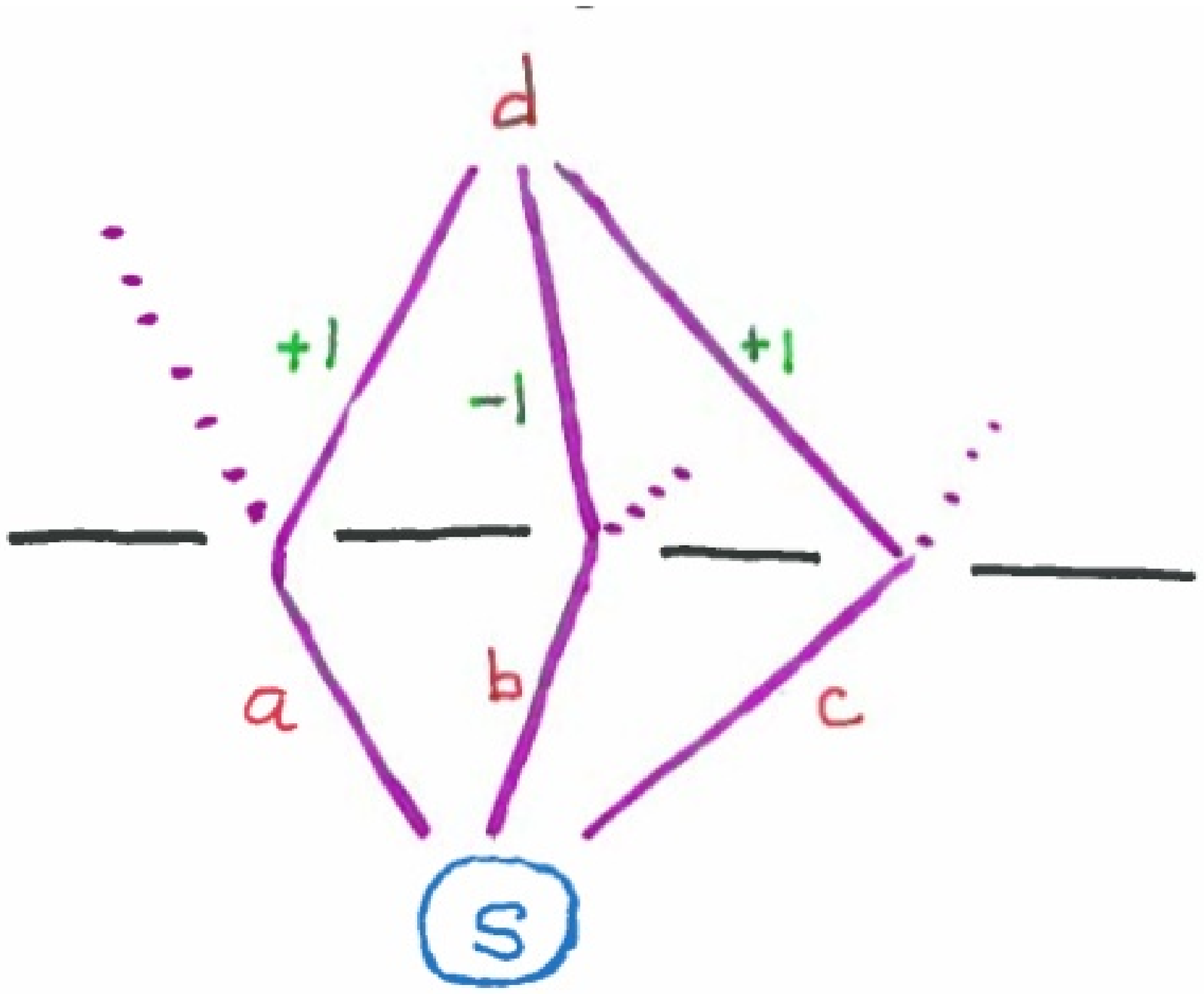}
{1}%
{The three-slit paradox}

To this setup belongs a history space $\Omega$ consisting of the various
possible particle trajectories, and a quantal measure $\mu$ assigning a
nonnegative real number to each set of trajectories.  
Let $a$ be the event that the trajectory passes through slit
$a$ and similarly for $b$ and $c$, and let $d$ be the event that it
arrives at $d$.  Consider further the event $A$ that the particle
arrives at $d$ after traversing $a$.  (Notice here that $a$, $b$ and $c$
are all intrinsic events, not measurement events.  We are not placing
detectors at any of the slits, either explicitly or implicitly.)
Writing the intersection $X\cap{Y}$ of two arbitrary events $X$ and $Y$
simply as their product $XY$, we have then that
$$
       A = ad, \ B=bd, \ C=cd, \ \ d=A+B+C \ ,
$$
where in the last equation a plus sign has been used to denote the
union of disjoint subsets. 

Now imagine the region $d$ to be small enough that we can represent the
path-integrals for $A$, $B$, and $C$ by single amplitudes whose squares
yield the (un-normalized) measures of the corresponding events, and
suppose further that these amplitudes  are $+1$ for events $A$ and
$C$ and $-1$ for $B$.  Then 
$\mu(d)=\mu(A+B+C)=|1-1+1|^2=1$, whereas
$$
    \mu(A+B) = \mu(B+C) = |1-1|^2 = 0  \ .
$$
Therefore, the events $A+B=d(a+b)$ and $B+C$ are precluded even though 
$A+B+C$ is not and can sometimes happen.

If we think classically, this is an outright contradiction.  Suppose we
look for the particle at $d$ and find it there.  We can then infer that
since it didn't pass through $a$ or $b$ ($A+B$ being precluded) it must
have arrived via $c$.  But reasoning symmetrically, we can infer by
the same token that it must have arrived via $a$.  Obviously, the two
conclusions contradict each other.

In the language of coevents, we can
express the situation in formulas as
$$
    \phi(A+B)=0,\  \phi(C+B)=0,\  \phi(d)=\phi(A+B+C)=1,
$$
where the first two equations follow from the preclusion rule and the
third expresses that the particle did arrive at $d$.
Formally a contradiction can be derived from these three equations by
Boolean manipulations following the classical rules of inference.
If one asks which rules were used (see the Appendix), 
one comes up with the following
list, 
where $A$ and $B$ represent arbitrary events and
$\negate{A}=\Omega\less{A}$ is the complementary event to $A$.  
\item{}From $\phi(A)=\phi(B)=1$ conclude $\phi(AB)=1$.
\item{}From $\phi(A)\not= 1$ conclude $\phi(A)=0$.
\item{}From $\phi(A)=0$ conclude $\phi(\negate{A})=1$.
\item{}From $A\part B$ and $\phi(A)=1$ conclude $\phi(B)=1$.

These formal relationships are instructive, but one can also
see the root of the  inconsistency informally in a way that indicates how
one might think to escape it.  What the formal rules
really express is the ingrained belief that reality is described by a
{\it single} trajectory $\gamma$ such that an event $A$ happens (the
corresponding predicate is true) if and only if $\gamma\in{A}$.  We
might therefore be able to extricate ourselves from the contradiction if
reality were given not by a single trajectory 
but by some more subtle combination of trajectories,
for which --- in some sense --- both $A$ and $C$ could  happen
simultaneously, or alternatively for which $A+C$ could happen without either $A$
or $C$ happening separately.  We will see that the so called
``multiplicative scheme'' realizes the latter alternative.

\section{5.~Freeing the coevent}
Which came first, the history or the event?  To the extent that an event
is nothing but a set of histories, it might seem that the history came
first, and this would be in accord with the classical worldview, where
only a single history is in some sense actual.\footnote{$^\dagger$}
{I'm leaving aside here questions of ``temporality'': is the past
 ``actual'', or the past and the future, or only the present, or \dots?  I
 hope that the neglect of such questions will not unduly prejudice the
 rest of this discussion.}
On the other hand when we consult our experience, what we meet with are
events.  Individual histories we experience --- if at all --- only as 
idealized limits of events.  One might argue on this basis that it is
the event that should come first, and this would be consistent with a
more ``holistic'' or ``dialectical'' attitude toward the history space
$\Omega$ (cf. category theory, toposes, etc.)

Be that as it may, the practical needs of probabilistic
theories, I think, force us to accord events an independent status, for
it is only they which have nontrivial measures in general.  For quantal
measures this argument becomes more convincing, because the measure of
an event no longer reduces, even formally to the measures of its
constituent histories.  (At best, it reduces to the measures of pairs
of histories.)  It seems clear in particular that the concept of
preclusion makes no sense at all except in relation to events.
Once events are dignified in this manner, the rules governing coevents
also acquire a certain freedom, and I am proposing to use this freedom
in order to overcome the logical conflict between preclusion and the
doctrine that reality can be described fully by a single history.  
More concretely, I am proposing to describe reality, not by an
individual history but by an individual coevent, which mathematically is
a kind of ``polynomial in histories''.  The rules governing which
coevents are dynamically possible can then change in such a way as to
accommodate the preclusion principle without engendering an
inconsistency.  
In the simplest case the polynomial is just a monomial, meaning in
effect just a {\it subset} $S$ of the history space.  This simplest
case, that of the {\it multiplicative scheme}, is the only one I will
discuss herein.  Other schemes are described in [16], 
[17] and [18].

I am tempted at this point just to present the multiplicative scheme and
discuss some of its applications, but I'm afraid that without further
background, it would appear far less natural than it will if its
intimate connection with deductive logic is brought out.
On the other hand, I know that for some people, any hint of tampering
with classical logic raises a barricade between them and the slightest
sympathy with whatever comes next.  For them I should emphasize that the
type of scheme I am proposing can stand as a self-contained framework,
whether or not one accepts a logical way of looking at it.
With this caveat, let me embark on some remarks relating logic to
physics that will lead in a natural way to the multiplicative scheme in
the next section.

For a scientist, logical inference is --- or I believe should be --- a
special case of dynamics.  Think for example of forecasting the motion
of mars using Kepler's laws.  Here we begin with certain events, the
locations of mars at certain earlier times, and from them we infer
certain other events, namely its locations at certain later times.  In
other cases, we may draw conclusions from the non-occurrence of an
event.  Thus, from the fact that the event ``sighting of the new moon''
did not happen last night, we might conclude that it will happen
tomorrow night (and in consequence a new month of the Islamic calendar
will begin).  This second example illustrates, I hope, how inferences
from dynamical laws can shade gradually into inferences from logic
alone, for example the inference that if my keys are not in my pocket
they must be in my jacket (which I left locked in my house).  In the
extreme case of simple abstract deductions like ``if `A' is true and `B'
is true then `A and B' is also true'', the inference feels so obvious
that we hardly recognize it as an inference at all, but this feeling
goes away for more complicated rules like ``Peirce's law''.\footnote{$^\flat$}
{$((A\to B) \to A) \to A$ }
Or think of the logical puzzles like,  ``the green house is to the right of
the white house, coffee is drunk in the green house,\dots''.

Notice here that the logic I'm speaking of concerns physical events,
not strings of words and not propositions in a formal language.  It is a
``logic of nature'', not a logic of language or
thought or mathematical truth.  This logic, I contend, is not
prior to experience.  Rather it codifies certain relations among events
that, 
until recently,
have been so ubiquitous in human experience
that they have been ossified and condensed
into a scheme that 
seems as if it were unchangeable.
What we do when we predict where mars
will appear next week is exactly what the rules of logical inference do
in a limited way, but also in an absolute and universal manner
reminiscent of how geometry was once treated as prior to physics.  But
just as a better understanding of gravity forced geometry back into
touch with physics, so also a better understanding of the microworld can
do the same for logic.  The resulting inferential scheme will not be
universal but will depend (at least in part) on the particular physical
system to which it is adapted.  This, at any rate, will be true for the
type of logic exemplified by the multiplicative scheme.

For present purposes, it helps to view a logic as a threefold
structure or ``triad'', 
whose outlines tend to be obscured when
formal logicians write about their subject.  The first component of the
triad is the event algebra $\EA$, 
a Boolean algebra
whose members can be thought of as
questions about the world, as described above.  If we adopt this imagery
then the second component of the triad is the space $\Z2$ of possible
answers (or ``truth values''), and the third (and most neglected) is the
``answering map'' or coevent $\phi:\EA\to\Z2$.  
In any given physical theory, 
$\EA$ and $\Z2$ will be fixed but $\phi$ will vary in the same
way that the solutions to Maxwell's equations vary.  Each
dynamically allowed $\phi$ describes then, a possible reality (or
``quantal history'' as one might term it).
In this context, dynamical ``laws of motion'' in the traditional sense and rules of logical
inference
both take the form of conditions on $\phi$.  
The
classical rules of inference can be stated very simply if we view them
in this manner.  In fact, we can give them in three equivalent forms,
one ``deductive'', one ``algebraic'' and one with a
topological or order-theoretic flavor.\footnote{$^\star$}
{I have not brought the quantifiers $\forall$ and $\exists$ into the
 discussion because they seem to be irrelevant.  One will implicitly use
 them in {formulating} the questions in $\EA$, but not in inferring
 relations among the possible answers.}

Before stating these rules however, I ought to highlight another aspect
of anhomomorphic logic that 
removes it from the more traditional milieu.
Namely, it pays equal
attention to both ``poles'' of $\Z2$, 
both $1$ (happening, truth) and $0$ (not-happening, falsehood).
Whether you know that an event has happened or that it has not
(as with the moon-sighting), 
you have learned something from which consequences can be drawn.  
Perhaps the reasons why falsehood has nevertheless tended to be ignored
in favor of truth are first,
that most logicians are mathematicians interested in 
deducing theorems from other propositions taken to be true; 
and second that they implicitly or explicitly adopt the rule
that $A$ is true if and only if $\negate{A}$ is false.  
In the context of physics and physical events, this rule is not as
self-evident as it might appear to be in a mathematical context, because
answering ``no'' to the question ``is the particle here?'' need not
commit you to answering ``yes'' to the question ``is the particle over
there?''.
Precisely this distinction will feature prominently in the
multiplicative scheme.  There is also evidence that logicians in early
Buddhist times took note of it. [19].

What then are the classical rules of inference
expressed as conditions on $\phi$?  
(I will assume in advance that $\phi$ is preclusive.)
In deductive form they can, as shown by Fay Dowker [20], 
be condensed into
three conditions
(where the symbol $\implies$ indicates deducibility): 

\item{(1a)}{ $\phi(A)=\phi(A\to B)=1 \ \implies\  \phi(B)=1$ \ (``modus ponens'')} 
\item{(1b)}{$\phi(A)=0 \ \implies\  \phi(\negate A)=1$}
\item{(1c)}{$\phi(0)=0 $} 

\noparagraph In algebraic form they are the condition:
\item{(2)}{ $\phi$ is a homomorphism of unital Boolean algebras ,}

\noparagraph
which says equivalently that $\phi$ preserves \& ({\it and}) and
$\negate$ ({\it not}).
And expressed as a condition on the events $\phi^{-1}(1)$ {\it affirmed}
by $\phi$ they say simply that
\item{(3)} {$\phi^{-1}(1)$ is a maximal preclusive filter in $\Omega$} \ .

\noparagraph
Here, following the usual definitions, a {\it filter} is a nonempty
family $\Phi$ of events 
(elements of $\EA$, hence subsets of $\Omega$) 
closed under intersection and passing to supersets.
It is {\it preclusive} if it contains no precluded events 
(whence it cannot contain the empty subset $0$), 
and it is 
{\it maximal} if it cannot be enlarged without ceasing to be preclusive.  
Poetically expressed, such a $\phi$ ``maximizes being'': it affirms as
many events as it can, subject to fulfilling the other conditions.

In view of formulation (2), a classical coevent may be called
``homomorphic'', and a coevent that breaks any of the classical rules may
be called {\it anhomomorphic}.

\section{6.~The multiplicative scheme: an example of anhomomorphic coevents}
We have granted ourselves the freedom to change the rules (``of inference'')
governing coevents, but how to do so?  Numerous avenues open up, but of
the large number that have been explored by those of us working on the
question (for some of them see [16]), only a handful have
seemed promising, in the sense of permitting enough events to happen on
one hand, but restricting the coevents sufficiently to reproduce the
predictive apparatus of standard quantum theory on the other hand.
The current favorite seems to be the {\it multiplicative scheme}, which
not only is among the simplest to apply, but also represents perhaps the
mildest change to the classical rules.  The change is so mild, in fact,
that it is non-existent if we express the classical rules in the form
(3) of the previous section!  The difference then springs solely from
the different meaning of ``preclusive'', or rather (because its meaning
as such has not changed) from the new patterns of preclusion (patterns of
precluded events) that become possible under the influence of quantal
interference.  

More formally, let us make the following definitions.  Recall that a
coevent $\phi$ is {\it preclusive} when it honors the preclusion
principle, that is when $\phi(A)=0$ for every precluded event $A\in\EA$.
Call such a $\phi$ {\it primitive} when it follows whatever further
rules of inference we have set up.  The collection of all primitive
coevents, I will denote by $\Ahat$, since it is analogous in some ways
to the spectrum of the event algebra $\EA$.  The elements of $\Ahat$ are
then (the descriptions of) the dynamically allowed ``realities'' or
``possible worlds''.

To arrive at the multiplicative scheme, we can retain condition (3) word
for word as the definition of a {\it primitive preclusive multiplicative
coevent}, or for short just ``primitive coevent''.  Sorting out the
definitions then shows that rules (1) and (2) do not survive intact.
Of the first set, (1a) and (1c) survive but (1b)
does not.  Of condition (2), what survives is unitality and the
preservation of the {\it and} operation (this being the origin of the
name ``multiplicative'', since algebraically, \& is multiplication).
All of condition (3) survives, of course, but its meaning is probably
easier to grasp when it is expressed in ``dual'' form.

To formulate it this way let us define first a map from sets $F$ of
histories to coevents $\phi=F^*$ by specifying that 
$F^*(A)=1$ iff $A\part{F}$.  To put this in words, let's say that a
coevent $\phi$ {\it affirms} an event $A$ when $\phi(A)=1$ and {\it denies}
it when $\phi(A)=0$.  Our definition then says that $F^*$ affirms precisely
those events that contain $F$
(as illustrated in the diagram, where $F^*$ affirms $A$ but denies both $B$
and $C$).
When $\phi=F^*$, I will say that $F$ is
the {\it support} of $\phi$.  Now,  in the case where $\Omega$ is a
finite set (which we are always assuming herein), one can check that
a multiplicative coevent necessarily takes the form $F^*$ for some support
$F\part\Omega$.  The condition (3) for primitivity then says precisely
that $F$ is as {\it small} as possible consistent with $\phi$
remaining preclusive.  Given the definition of $F^*$, the condition for
primitivity thus boils down to a rather simple criterion:
the support should shrink down as much as possible without
withdrawing into any precluded event.

\epsfxsize=3.6in
\bigskip
\FigureNumberCaption
{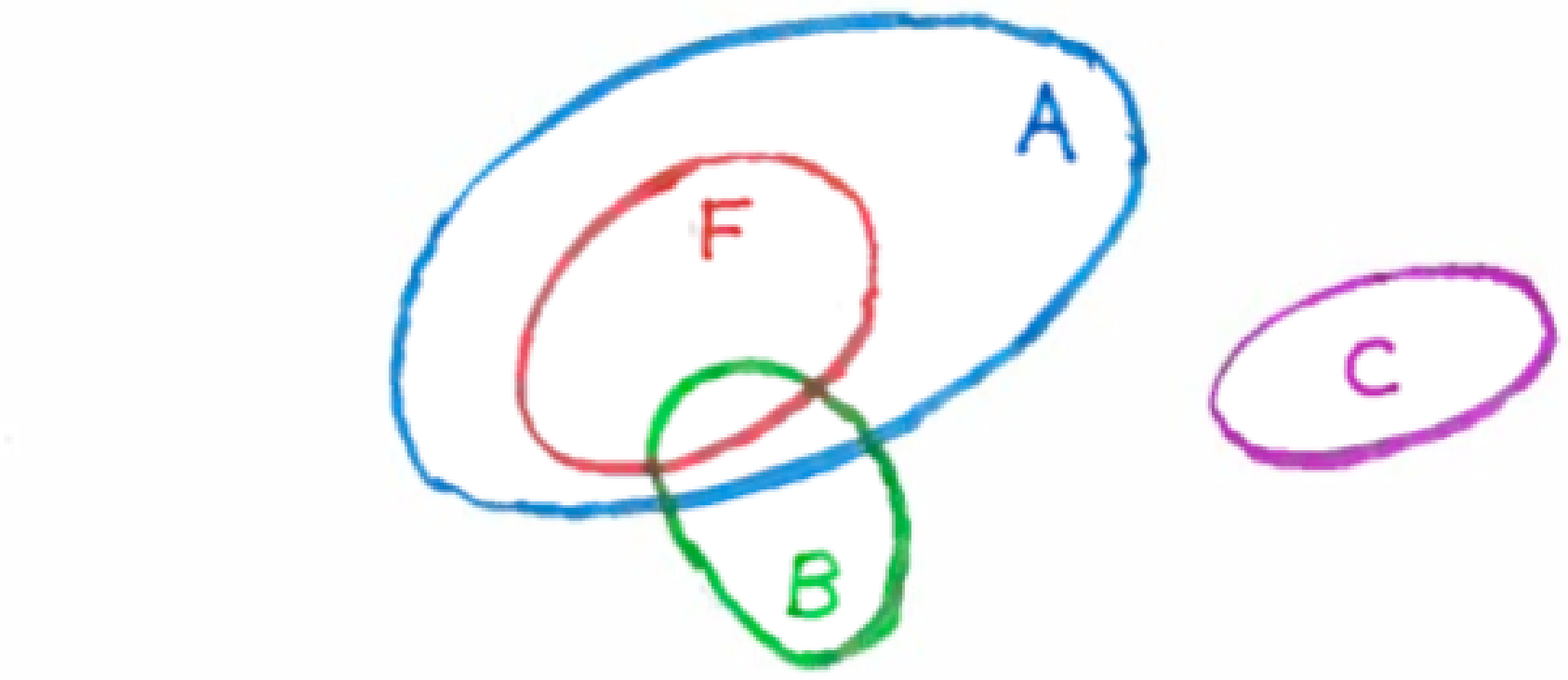}
{2}%
{Three events and the coevent $\phi=F^*$}

As remarked above, ``truth'' or ``happening'' is in this context a
``collective property'', since it pertains to events rather then to
individual histories.  A multiplicative coevent is also collective in
nature, since it corresponds to a subset of $\Omega$ rather than an
individual element.

A first test of any scheme of the present sort is that it should
reproduce the classical notion of reality 
(namely reality as a single history) 
when the pattern of preclusions is itself classical.
({We can take the latter to mean that an arbitrary event is precluded if and
 only if it is covered by precluded events.  In particular, every
 subevent of a precluded event must be precluded.})
In particular, this should happen when the quantal measure $\mu$ reduces
to an ordinary measure, and also in the case of deterministic theories
like classical mechanics where the dynamics reduces simply to the
preclusion of an entire class of histories --- those which fail to satisfy
the equations of motion.  It is not too difficult to verify that the
multiplicative scheme passes the test in both cases.  
(See the theorems in Section 7.)

\subsection{Resolution of the 3-slit paradox}
It is a feature of the multiplicative scheme that any event $E$ can find
a primitive coevent to affirm it, as long as it is not included in some
other event of zero measure.  That is, there will be at least one
{$\phi\in\Ahat$} such that $\phi(E)=1$.  In our three-slit example, the
events $A+C$ and $A+B+C=d$ are both of this type, so we can see already
that the multiplicative scheme will avoid the false prediction that $d$
can never occur.

To simplify things, let's imagine that there is nothing in existence but
this particular experiment and let us further ignore all histories not
in $d$ and all fine structure of  the histories that are in $d$.  Then
$\Omega=d$ consists of only three elements, identifiable with the three
``atoms'', $A$, $B$, and $C$, of the event algebra $\EA$.  With such a
small history space it is easy to enumerate all the possible 
(multiplicative) coevents,
and one sees by inspection that only two are preclusive, namely
$\phi=(A+C)^*$ and  $\phi=(A+B+C)^*$.  The latter, however, is not
primitive, since the former has smaller support. There is thus a unique
primitive coevent, $\phi=(A+C)^*=A^*C^*$.  With respect to this coevent, two
of the eight events in $\EA$ happen, namely $A+C$ and $d$ itself, and the other 
six do not, namely $A$, $B$, $C$, $0$, and of course $A+B$ and $B+C$
($0$ being the empty subset of $\Omega$).  In particular $\phi(d)=1$,
so the paradox is removed.

This is satisfactory as far as it goes, but in one respect this 3-slit
example is misleadingly simple.  As the cardinality $N$ of the history
space grows, it becomes increasingly difficult in practice to work out
the primitive coevents (in the multiplicative scheme there are $2^N$
potential supports to consider), but when the dynamics is simple enough
it is possible to count them or at least to estimate their number.
Typically one finds that this number also grows rapidly with $N$, just
as one might have expected.  The fact that {$\phi$} is unique for the
3-slit setup is thus very much of an exception.

There is also another respect in which our example has been overly
idealized.  We have cut the experiment off at the point where the
particle reaches (or does not reach, as the case may be) the location
$d$, thereby ignoring, not only the future, but also whatever else is
going on in the world besides this experiment.  Both of these omissions
could have serious repercussions which I'll return to briefly in the
concluding section.

\section{7.~Preclusive separability and the ``measurement problem''}
Within the framework we have arrived at, individual histories
are replaced in a certain sense by sets of histories while ``laws of
motion'' are expressed via preclusion and the requirement of
primitivity.  In this way dynamics merges with logic to some extent, and
we are able to speak directly about microscopic processes without
succumbing to paradoxes of the Kochen Specker sort --- at least in
simple examples.  Because of its ``realistic'' nature, I hope that this
framework will prove useful in connection with quantum gravity,
specifically in the quest for a causal set dynamics of quantal
sequential growth.  But a more immediate challenge is posed by the so
called ``measurement problem''.  If the multiplicative scheme cannot
solve this problem, it will be hard to take it seriously as a potential
basis for unifying quantum field theory with general relativity.

Of course there is no single, well posed ``measurement problem''.
Rather this phrase refers to a complex of issues concerning the
relationship of quantal processes to the macroscopic realm of classical
events and ``observers''.  Nevertheless, I think one would not be
oversimplifying unduly to pose the problem as that of accounting for
measurements without resorting to the notion of external agents who
are not explicable in microscopic terms.  In the context of the
multiplicative scheme (or any of the other schemes based on
anhomomorphic coevents) this problem acquires a precise formulation.
One must show that the primitive coevents become classical
(i.e. homomorphic) when they are restricted to a suitable subalgebra
$\EA^{macro}$ of ``instrument events''.

To appreciate that this is what is needed, recall why there is a problem
in the first place.  Quantum mechanics as ordinarily presented either
declines to describe the measurement process or it gives a manifestly
false description, depending on whether or not one assumes that the
state-vector ``collapses'' during the measurement.  In the former case
one is positing a phenomenon that the theory leaves in the dark, in the
latter case the theory serves up a superposition of macroscopically
distinct outcomes that contradicts our most elementary experiences.  
Now let us return to the coevent framework, where measurements are no
different in principle from other quantal processes (and like other
processes are to be described in terms of histories rather than evolving
state-vectors.)  In a measurement-like situation, the theory will yield
a definite set of primitive coevents to describe the different possible
outcomes.  For example let events $A$ and $B$ be two alternative
``pointer readings'' in some experiment.  Each of these events will
correspond to a particular collection of configurations of the ``pointer
molecule worldlines'', and will be macroscopic in the sense that the
corresponding histories will
involve large numbers of particles, relatively great masses, etc.  If a
given coevent $\phi\in\Ahat$ affirms $A$ and denies $B$ then $A$ is the
outcome in the world described by $\phi$, in the contrary case it is
$B$.  (Both types of coevent can be viable in general, since the theory
is not deterministic.)\footnote{$^\dagger$}
{Having written this however, I should add that in simple examples, the
 theory turns out to be much closer to deterministic than one might have
 expected.  (See the next section.) }
%
%
However, one can also construct ``Schr{\"o}dinger cat''-like coevents
which deny both $A$ and $B$, as in the 3-slit example above.  Such a
coevent would not be in accord with experience, which always (or almost
always?) presents us with a unique outcome that does happen.
Consistency with experience thus requires that no (or almost no) coevent
$\phi\in\Ahat$ be of this ambiguous type, and this in turn is equivalent to
$\phi|\EA^{macro}$ (the restriction of $\phi$ to $\EA^{macro}$) being
classical, since when classical logic reigns, precisely one history
occurs.

Formally, we can define a subalgebra $\EA^{macro}\part\EA$ of
macroscopic events\footnote{$^\flat$}
{For the events comprising $\EA^{macro}$ to be well defined, we might
 have first to condition on the happening of certain other events that
 are prerequisite to the existence of macroscopic objects, i.e. to the
 existence of what is sometimes called a ``quasiclassical realm''.}
such that disjoint elements $A$ and $B$ of $\EA^{macro}$ correspond to
macroscopically distinct events in $\EA$.   
Such a subalgebra induces a partition of $\Omega$ whose equivalence
classes (sets of histories distinguished by no element of $\EA^{macro}$) define
a quotient or ``coarse-graining'' $\Omega^{macro}$ of $\Omega$ into
``macroscopic histories''.  Our condition that $\phi$ map $\EA^{macro}$
homomorphically into $\Z2$ is then trying to say that $\phi$ is
supported within a single coarse-grained history (the translation being
literally correct when $|\Omega|<\infty$ ).
Put differently, the support $F$ of such a $\phi$ must not overlap
macroscopically distinct events.  When this condition is satisfied $F^*$
will look to $\EA^{macro}$ like a single coarse-grained history
and will be classical in that sense.

Given the measurement problem rendered in this manner, we can solve it
if we can find a sufficient condition for $\phi$ to behave classically
and if in addition we can give reasons why the events of our macroscopic
experience (almost)
always fulfill the condition.  To illustrate how this can work, I will
quote two theorems that furnish sufficient conditions of the type we
need.  It should be clear from the preceding discussion that when either
theorem applies, primitive coevents will behave classically as far as
instrument events are concerned.  One can also see the same thing by
translating the conclusion of the theorems into the statement that rule
(1b) above is respected.  Since the multiplicative scheme validates the
rest of rule (1) by construction, (1b) suffices to return us to the
classical case.

The first theorem below furnishes a sufficient condition that is easy to
state but more restrictive than it needs to be.  The condition in the
second theorem is less transparent in statement, but arguably more
likely to hold in practice.  
(A proof of the first theorem in the finite case can be found in
[16].
Proofs of the second theorem and the infinite case of the
first theorem exist as well, but remain unpublished. )

\THEOREM 1.  \ 
     Let $\Omega = \Omega' + \Omega''$ 
     be a partition of $\Omega$ such that an arbitrary event
     $A\part\Omega$ is precluded iff its intersections with 
     $\Omega'$ and $\Omega''$ are both precluded, and 
     let $\phi$ be any primitive preclusive coevent in the
     multiplicative scheme.
     Then $\phi$ is supported within either $\Omega'$ or $\Omega''$.
     That is, $\phi=F^*$ with $F\part\Omega'$ or $F\part\Omega''$.

\THEOREM 2.  \ 
   The conclusion of theorem 1 persists if both $\Omega'$ and $\Omega''$
   satisfy the following weaker condition on subsets $S$ of $\Omega$:
       If any event $A\part S$ lies within any precluded event $B$ at all
       then it lies within a precluded event $C\part S$.

As we have seen, the important consequence of these theorems is that,
when either of the respective conditions is satisfied, and with respect
to any primitive coevent $\phi$, either $\Omega'$ or $\Omega''$ happens but
not both.
The important question then becomes whether macroscopic events are in
fact ``preclusively separable'' in this way.  This would follow
immediately from the still stronger condition that: No event in
$\Omega'$ interferes with any event in $\Omega''$.  However this condition
represents a very strict type of ``decoherence'' closely related to idea
of a {\it record}.\footnote{$^\star$}
{In the context of unitary quantum mechanics, the condition is satisfied
 for records because different versions of a given record correspond to
 disjoint regions in configuration space.  The weaker type of decoherence
 usually contemplated by ``decoherent historians''
 requires only that $\Omega'$ and $\Omega''$ (belong to $\EA$ and)
 decohere, not that their arbitrary subevents decohere as well.}
To the extent that one is willing to posit the existence of sufficiently
permanent records of macroscopic events, one can therefore regard the
measurement problem as solved.  To the extent that one finds this
assumption implausibly strong one can still hope to prove that
macroscopic events fulfill the conditions of one of the theorems.  In
this sense, the measurement problem reduces to a calculation.

\section{8.~Open questions and further work}
Before the framework presented above can be considered complete, further
work will be needed on some of the questions raised by the above
discussion.  Foremost among these is probably the question whether one
can demonstrate by examples or general arguments that the events of our
macroscopic experience
really are preclusively separable in the sense 
of the above theorems. 
Assuming that they are, can one explain on this
basis why the textbook paradigm involving the wave function and its
``collapse'' works as well as it does, and if so can one quantify the
deviations that one should expect from this paradigm?  Here much of the
way forward is clear.  There exists a sketch of a derivation of the collapse
rule, but it needs to be followed out in more detail.  In the
same direction, we of course need to recover Born's rule for
probabilities, either by appeal to 
approximate preclusion and 
the Cournot principle
(cf. [11]) or in some better manner.  
And finally, the definition of primitive coevent 
needs to be extended to infinite event algebras, since the most
important examples of quantal dynamics (atomic physics, quantum field
theory, etc.) are of this type, at least in current idealizations.  Here
again, there is much that could be said about work already done.

Even in its partly finished state, the coevent framework, like the
Bohmian version of quantum mechanics, lets us pose questions that we
would not have been able to formulate from a ``Copenhagen'' standpoint.
Thus for example, one can ask for the primitive coevents that describe
the ground state of a Hydrogen atom, or of a particle in a harmonic
oscillator potential.  The Bohmian particle in these cases just sits
still wherever it happens to find itself.  One wouldn't know by
following its motion what kind of force was binding it, nor could
one even know its energy in many cases.  It's therefore of particular
interest to ask of the multiplicative scheme what sets of trajectories
comprise the supports of the primitive coevents in these cases.  Could
one deduce from these sets of trajectories what the potential was and
would the energy show up clearly?  Currently such questions seem nearly
beyond reach, in the first place because of the mathematical
difficulties in defining the continuum path integral itself on a sufficiently
large domain of events. (cf. [21])

More accessible, though no less interesting, are questions about
experiments of the Kochen-Specker type, or about entangled pairs of
particles passing through successive Stern-Gerlach analyzers.  For 
a few gedankenexperiments of this type people have been able to find
some or all of the primitive coevents, and in some cases to study causal
relationships between the coevents at earlier and later stages of the
process [22] [23].
Such examples can serve as laboratories to explore
possible meanings for relativistic causality, locality, and determinism
within the coevent framework.  For example, in a simple extension of the
Hardy experiment, one finds 286 coevents $\phi$ of which 280 behave
deterministically in the sense that the restriction of $\phi$ to the
subalgebra of past events uniquely determines $\phi$ globally
(cp. a similar effect found in [24]).
One can also formulate different conditions of relativistic causality
(``screening off'') for such systems and study to what extent, and in
what circumstances they hold.  If a suitable condition could be found,
it could then be carried over to the causal set situation and used as a
guide to formulating a quantal analog of the {classical} sequential
growth models (cf. [14]).

If one regards the coevent schemes in logical terms, it's natural to try
to bring them into relation with other non-classical logics to which
anhomomorphic inference seems to bear some resemblance, such as
intuitionistic, dialectical or paraconsistent logic.  With dialectics,
anhomomorphic logic shares a certain tolerance of contradiction, or of
what classically would be regarded as self contradictory.  With
intuitionism it shares a non-classical understanding of negation, 
not however at
the level of the event algebra, which remains strictly Boolean so that
$\negate\negate{A}=A$ for all $A\in\EA$, but at the level of inference,
where $\phi(A)$ becomes independent of $\phi(\negate{A})$.  On the other
hand, whereas intuitionistic logic simply drops certain rules of
inference like proof by contradiction, anhomomorphic logic adds
crucially the new requirement of {\it primitivity}.  Thus it cannot be
characterized simply as either weaker or stronger than classical logic.

Returning to the coevent framework per se, I'd like to allude briefly
some of its more radical consequences, and the risks (or possibly
opportunities) they hold out for this way of conceiving quantal reality.
Each of these consequences is visible in simple examples.  In the Hardy
example referred to just above, one potentially encounters what could be
called ``premonitions''.  To be confident of their occurrence one would
have to incorporate ``instrument setting events'' into the model, and
this has not been done.  Yet it looks as if the past of the coevent can
determine not only events involving the particles in question, but also
to some extent the settings themselves.  Such an effect could be called
a premonition on the part of the particles, but it could also be called
a cause of the later setting-event, in which case there would be no
suggestion of ``retro-causality''.  In the simple example of a particle
hopping unitarily between the nodes of a two-site lattice (``two site
hopper''), one encounters a potential danger that also shows up in much
the same form in connection with composite systems.  In both cases it
can happen that the restriction of the coevent to the subalgebra of
early-time events (respectively events in one of the two subsystems)
trivializes in the sense that its support becomes the whole space of
partial histories.  This means that only correlations between early and
late times (resp. between one subsystem and the other) happen
nontrivially.  None of this is a problem unless carried to extremes.  If
for example, the relevant time scale for this type of trivialization in
realistic systems were to be comparable with the Poincar{\'e} recurrence
time, then there would be little to worry about.

Finally, let me conclude with a possibility that for now is merely a
dream, but which, if it came to pass, would bring with it a striking
historical irony.  One might discover laws that governed the pattern of
preclusions without referring, directly or indirectly, to the quantal
measure $\mu$.  If that happened, it would provide a more radical
revision of classical dynamics (stochastic or deterministic) than that
represented by the path integral.  
Or in the process of working out the
primitive coevents in various examples, one might even discover laws
expressed directly for the coevents themselves, without even needing to
derive them from preclusions.  
If that happened, 
the whole superstructure of amplitudes and generalized measures would
fall away, and
quantum theory would
have led back to something resembling classical 
equations of motion, 
but at
a higher ``structural'' level than occupied by our old theories that
identified reality with a single history.



\bigskip
\noindent

The ideas presented above have grown out of extensive joint work with 
Fay Dowker, 
Cohl Furey, 
Yousef Ghazi-Tabatabai, 
Joe Henson, 
David Rideout, 
and
Petros Wallden.
Research at Perimeter Institute for Theoretical Physics is supported in
part by the Government of Canada through NSERC and by the Province of
Ontario through MRI.
%

\section{Appendix.~Formal deduction of the three-slit contradiction}
We are given the preclusions $\phi((a+b)d) = \phi((b+c)d) = 0$ and wish
to deduce that $\phi(d)=0$, assuming that $\phi$ follows the classical
rules of inference. 

     {step 0:} Suppose that $\phi(d)=1$.

     {step 1:} if also $\phi(a+b)=1$ then $\phi((a+b)d)=1$ 
             contrary to what was given, \lbr 
             \phantom{step 400} hence $\phi(a+b)=0$.

     {step 2:} $\phi(b+c) = 0$ by symmetry.

     {step 3:} $\phi(a+b)=0$ implies $\phi(c)=1$ since $c=\negate(a+b)$,
     the complement of $a+b$.

     {step 4:} From $\phi(c)=1$ follows $\phi(b+c)=1$, contradicting step 2.

     {step 5:} Therefore our supposition was false and $\phi(d)=0$.

\noparagraph What conditions on $\phi$ did we use?

      In step 1: If $\phi(A)=\phi(B)=1$ then $\phi(AB)=1$; if
      $\phi(A)\not=1$ then $\phi(A)=0$.

      In step 3: If $\phi(A)=0$ then $\phi(\negate{A})=1$

      In step 4: If $A\part B$ and $\phi(A)=1$ then $\phi(B)=1$


\noparagraph  In the multiplicative scheme, only step 3 would fail.
Notice that in reasoning {\it about} $\phi$ we have also employed
classical logic, in particular proof by contradiction in steps 1 and 5.

\ReferencesBegin                             

\ref [1]  
Luca Bombelli, Joohan Lee, David Meyer and Rafael D.~Sorkin, 
``Spacetime as a Causal Set'', 
  \journaldata {Phys. Rev. Lett.}{59}{521-524}{1987};
\sepref
Rafael D.~Sorkin,
``Causal Sets: Discrete Gravity (Notes for the Valdivia Summer School)'',
in {\it Lectures on Quantum Gravity}
(Series of the Centro De Estudios Cient{\'\i}ficos),
proceedings of the Valdivia Summer School, 
held January 2002 in Valdivia, Chile, 
edited by Andr{\'e}s Gomberoff and Don Marolf 
(Plenum, 2005) \lbr
\arxiv{gr-qc/0309009};
\sepref
Joe Henson, ``The causal set approach to quantum gravity''
 \eprint{gr-qc/0601121}
 This is an extended version of a review to be published in 
  ``Approaches to Quantum Gravity - Towards a new understanding of space and time'' 
 (ed. D. Oriti), 
 (Cambridge University Press, 2006).

\ref [2] David P.~Rideout and Rafael D.~Sorkin,
``A Classical Sequential Growth Dynamics for Causal Sets'',
 \journaldata{Phys. Rev.~D}{61}{024002}{2000} \lbr
 \arxiv{gr-qc/9904062}

\ref [3] Rafael D. Sorkin, ``Quantum dynamics without the wave function''
  \journaldata{J. Phys. A: Math. Theor.}{40}{3207-3221}{2007}
  (http://stacks.iop.org/1751-8121/40/3207)
  \eprint{quant-ph/0610204} \lbr
  \hpf{http://www.physics.syr.edu/~sorkin/some.papers/}

\ref [4] Rafael D.~Sorkin, ``Quantum Mechanics as Quantum Measure Theory'',
   \journaldata{Mod. Phys. Lett.~A}{9 {\rm (No.~33)}}{3119-3127}{1994}
   \eprint{gr-qc/9401003}\lbr
   \eprint{http://www.perimeterinstitute.ca/personal/rsorkin/some.papers/}

\ref [5] Roberto B.~Salgado, ``Some Identities for the Quantum Measure and its Generalizations'',
 \journaldata{Mod. Phys. Lett.}{A17}{711-728}{2002}
 \eprint{gr-qc/9903015}

\ref[6] Rafael D.~Sorkin, ``Quantum Measure Theory and its Interpretation'', 
  in
   {\it Quantum Classical Correspondence:  Proceedings of the $4^{\rm th}$ 
    Drexel Symposium on Quantum Nonintegrability},
     held Philadelphia, September 8-11, 1994,
    edited by D.H.~Feng and B-L~Hu, 
    pages 229--251
    (International Press, Cambridge Mass. 1997)
    \eprint{gr-qc/9507057}
  \hpf{http://www.physics.syr.edu/~sorkin/some.papers/}

\ref [7] Lucien Hardy, ``Quantum Theory From Five Reasonable Axioms'', \lbr
 \arxiv{quant-ph/0101012v4}

\ref [8] J.B.~Hartle, ``Spacetime Quantum Mechanics and the Quantum Mechanics of Spacetime'',
 in B.~Julia and J.~Zinn-Justin (eds.),
 {\it Gravitation et Quantifications: Les Houches Summer School, session LVII, 1992}
 (Elsevier Science B.V. 1995) \lbr
 \arxiv{gr-qc/9304006}

\ref [9] Urbasi Sinha, Christophe Couteau, Zachari Medendorp, Immo S{\"o}llner,
Raymond La\-flamme, Rafael D. Sorkin, and Gregor Weihs,
``Testing Born's Rule in Quantum Mechanics with a Triple Slit Experiment'',
    in {\it Foundations of Probability and Physics-5}, 
    edited by L. Accardi, G. Adenier, C. Fuchs,
    G. Jaeger, A. Yu. Khrennikov, J.\AA. Larsson, S. Stenholm, 
    American Institute of Physics Conference Proceedings,
    Vol. 1101, pp. 200-207 (New-York 2009)
  (e-print: arXiv:0811.2068 [quant-ph])
 %

\ref [10] Yousef Ghazi-Tabatabai and Petros Wallden, ``Dynamics \& Predictions in the Co-Event Interpretation'',
 Journal-ref: J. Phys. A: Math. Theor. 42 (2009), 235303 
 \arxiv{0901.3675}

\ref [11] A.N. Kolmogorov, {\it Grundbegriffe der Wahrscheinlichkeitsrechnung, von A. Kolmogoroff}
(Berlin, Springer 1933).
translated as: \sepref
A.N. Kolmogorov, {\it Foundations of the Theory of Probability}
(New York, Chelsea Pub. Co. 1956)

\ref [12] Allen Stairs,
\journaldata{Phil. Sci.}{50}{578}{1983}; and private communication

\ref [13] Sukanya Sinha and Rafael~D.~Sorkin, ``A Sum-over-histories Account of an EPR(B) Experiment'',
   \journaldata{Found. of Phys. Lett.}{4}{303-335}{1991}

\ref [14] David Craig, Fay Dowker, Joe Henson, Seth Major, David Rideout and Rafael D.~Sorkin,
``A Bell Inequality Analog in Quantum Measure Theory'',
\journaldata{J. Phys. A: Math. Theor.}{40}{501-523}{2007},
\eprint{quant-ph/0605008}, \lbr
\eprint{http://www.perimeterinstitute.ca/personal/rsorkin/some.papers/}

\ref [15] Sumati Surya and Petros Wallden, ``Quantum Covers in Quantum Measure Theory'',
\arxiv{0809.1951}

\ref [16] Yousef Ghazi-Tabatabai,
{\it Quantum Measure Theory: A New Interpretation} \lbr
\arxiv{0906.0294 (quant-ph)}

\ref [17] Rafael D. Sorkin, ``An exercise in `anhomomorphic logic'~'',
 \journaldata{Journal of Physics: Conference Series (JPCS)}{67}{012018}{2007},
 a special volume edited by L. Diosi, H-T Elze, and G. Vitiello, and
 devoted to the Proceedings of the DICE-2006 meeting,
  held September 2006, in Piombino, Italia.
  \eprint{arxiv quant-ph/0703276} , \lbr
\hpf{http://www.physics.syr.edu/~sorkin/some.papers/}

\ref [18] Stan Gudder, ``An anhomomorphic logic
 for quantum mechanics'' \lbr
\arxiv{0910.3253 (quant-ph)}

\ref  [19] Rafael D.~Sorkin, ``To What Type of Logic Does the ``Tetralemma'' Belong?'', \lbr
\arxiv{10035735 (math:logic)} \lbr
\hpf{ http://www.physics.syr.edu/~sorkin/some.papers/ }

\ref [20] Fay Dowker and Petros Wallden,
``Modus Ponens and the Interpretation of Quantum Mechanics''
(in preparation)

\ref [21] Stan Gudder, ``Quantum measure and integration theory'',
\journaldata{J. Math. Phys.}{50}{123509}{2009}
\arxiv{0909.2203 (quant-ph)}

\ref [22] Fay Dowker and Yousef Ghazi-Tabatabai, ``The Kochen-Specker Theorem Revisited in Quantum Measure Theory'',
 \journaldata {J.Phys.A}{41}{105301}{2008} \lbr
 \arxiv{0711.0894 (quant-ph)}

\ref  [23] Cohl Furey and Rafael D.~Sorkin,
``Anhomomorphic Co-events and the Hardy Thought Experiment''
(in preparation)

\ref [24] Fay Dowker and Isabelle Herbauts, ``Simulating causal collapse models'',
\journaldata{Classical and Quantum Gravity}{21}{2963-2980}{2004} 
\arxiv{quant-ph/0401075}

\ref [25] Yousef Ghazi-Tabatabai and Petros Wallden, ``The emergence of probabilities in anhomomorphic logic''
 \journaldata{Journal of Physics: Conf. Ser.} {174} {012054} {2009} \lbr
 \arxiv{0907.0754 (quant-ph)}

\end


(prog1    'now-outlining
  (Outline 
     "\f......"
      "
      "
      "
   ;; "\\\\message"
   "\\\\Abstrac"
   "\\\\section"
   "\\\\subsectio"
   "\\\\appendi"
   "\\\\Referen"
   "\\\\ref....[^|]"
  ;"\\\\ref....."
   "\\\\end